\documentclass[journal, 10pt]{IEEEtran}
\IEEEoverridecommandlockouts
\usepackage{amsmath}
\usepackage{cite}
\usepackage{amsmath,amssymb,amsfonts}
\usepackage[ruled,linesnumbered,lined]{algorithm2e}
\usepackage{graphicx}
\usepackage{textcomp}
\usepackage{xcolor}
\usepackage{bm}
\usepackage{booktabs}
\usepackage{caption,subcaption}
\usepackage{amsthm,amssymb,amsfonts}

\newtheorem{lemma}{Lemma}

\SetKwInput{Input}{input}
\SetKwInput{Output}{output}

\usepackage{array}
\newcolumntype{L}[1]{>{\raggedright\let\newline\\\arraybackslash\hspace{0pt}}m{#1}}
\newcolumntype{C}[1]{>{\centering\let\newline\\\arraybackslash\hspace{0pt}}m{#1}}
\newcolumntype{R}[1]{>{\raggedleft\let\newline\\\arraybackslash\hspace{0pt}}m{#1}}

\newtheorem{proposition}{Proposition}

\newcommand{\rmj}{{\mathrm j}}

\newcommand{\rmt}{{\mathrm t}}
\def\BibTeX{{\rm B\kern-.05em{\sc i\kern-.025em b}\kern-.08em
		T\kern-.1667em\lower.7ex\hbox{E}\kern-.125emX}}

\begin{document}
	\title{On the Performance of Lossless Reciprocal MiLAC Architectures in Multi-User Networks}
     \author{Tianyu Fang, \IEEEmembership{Graduate Student Member, IEEE}, Xiaohua Zhou, \IEEEmembership{Graduate Student Member, IEEE},  and Yijie Mao, \IEEEmembership{Member, IEEE}
\vspace{-0.8cm}
\thanks{
\par T. Fang is with the Centre for Wireless Communications, University of Oulu, 90014 Oulu, Finland (e-mail:
tianyu.fang@oulu.fi). 
\par X. Zhou and Y. Mao are with the School of Information Science and Technology, ShanghaiTech University, Shanghai 201210, China (e-mail:
			{zhouxh3, maoyj}@shanghaitech.edu.cn).
            }}

	\maketitle
\vspace{-1.2cm}
	\begin{abstract}
		Microwave linear analog computer (MiLAC)-aided beamforming, which processes the transmitted symbols fully in the analog domain, has recently emerged as a promising alternative to fully digital and hybrid beamforming architectures for multiple-input multiple-output (MIMO) systems. While prior studies have shown that lossless and reciprocal MiLAC can achieve the same capacity as digital beamforming in a single-user MIMO network, its performance in multi-user scenarios remains unknown. 
        To answer this question, in this work, we establish a downlink multi-user multiple-input single-output (MU-MISO) network with a MiLAC-aided transmitter, and investigate its sum-rate performance. Based on the microwave network theory, we first prove that  lossless and reciprocal MiLAC cannot achieve the same performance as digital beamforming in a general MU-MISO network. Then, we formulate a sum-rate maximization problem and  develop an efficient optimization framework to jointly optimize the power allocation and the scattering matrix for MiLAC. Numerical results validate our theoretical analysis and demonstrate that MiLAC is a promising architecture for future extremely large-scale MIMO systems.

  \end{abstract}
	
	\section{Introduction}
    Microwave linear analog computer (MiLAC) has recently been proposed as a promising solution for extremely large-scale MIMO systems~\cite{nerini2025analog,nerini2025enabling}.
    Utilizing a reconfigurable multiport microwave network composed of tunable impedance components, MiLAC directly manipulates signals in the electromagnetic (EM) domain. Specifically, the transmitted signal enters the input ports, is shaped  via analog microwave interactions as it propagates through the network, and exits at the output ports for direct antenna radiation. 
    By leveraging this inherent analog computation ability, we can use a MiLAC at the transmitter to perform analog beamforming. In this setup, baseband symbols are first converted to radio-frequency (RF) signals and provided to the MiLAC input ports through RF chains, while its output ports are connected to the transmit antenna arrays. 
    In this way, the MiLAC enables efficient beamforming without relying on high-dimensional digital or hybrid precoding circuits, offering a compact and energy-efficient solution for next-generation wireless systems.
    
   MiLAC-based beamforming has been shown in \cite{nerini2025enabling} that, with ideal impedance tuning, it achieves the same performance as digital zero-forcing (ZF) beamforming at just $1/15000$th of the computational cost. To move toward a more realistic design, \cite{nerini2025capacity} examined a lossless and reciprocal MiLAC architecture with purely imaginary and symmetric impedances, and proved that it achieves the capacity of a single-user multiple-input multiple-output (SU-MIMO) system. 
   However, this design, known as the fully-connected MiLAC, requires all ports to be interconnected. It provides maximum flexibility at the cost of substantial circuit complexity and control overhead. 
   To address this limitation, \cite{nerini2025mimo} proposed stem-connected MiLACs, a family of architectures that preserve capacity-achieving performance in SU-MIMO while significantly reducing circuit complexity. 
   While existing works \cite{nerini2025capacity,nerini2025mimo} have demonstrated that lossless reciprocal MiLAC can achieve the capacity of a SU-MIMO system, their performance in general multi-user scenarios remains unknown. This leads to a core research question:  \emph{Is MiLAC a capacity-achieving strategy in a general multi-user multi-antenna network?}

In this work, we address this open question by investigating the performance a downlink multi-user multiple-input single-output (MU-MISO) network with a fully-connected, lossless, and reciprocal MiLAC at the transmitter. 
We first establish the relationship between the MiLAC transmit beamforming  and  scattering matrices. Using this result, we characterize the achievable performance upper bound for MiLAC-based beamforming, and  prove analytically that a lossless and reciprocal MiLAC cannot attain the same performance as digital beamforming in a general MU-MISO network. Then, we  formulate a sum-rate maximization problem for MiLAC, and develop an efficient optimization framework that iteratively updates the RF-chain power allocation and the MiLAC scattering matrix with closed-form solutions. Numerical results validate our theoretical analysis and demonstrate the effectiveness of the proposed algorithm.

\section{System  Model and Problem Formulation}\label{Sec:system model}
	\subsection{System Model}\label{sec:SysMod}
	
\begin{figure}[t]
		\centering
		\includegraphics[width=0.48\textwidth]{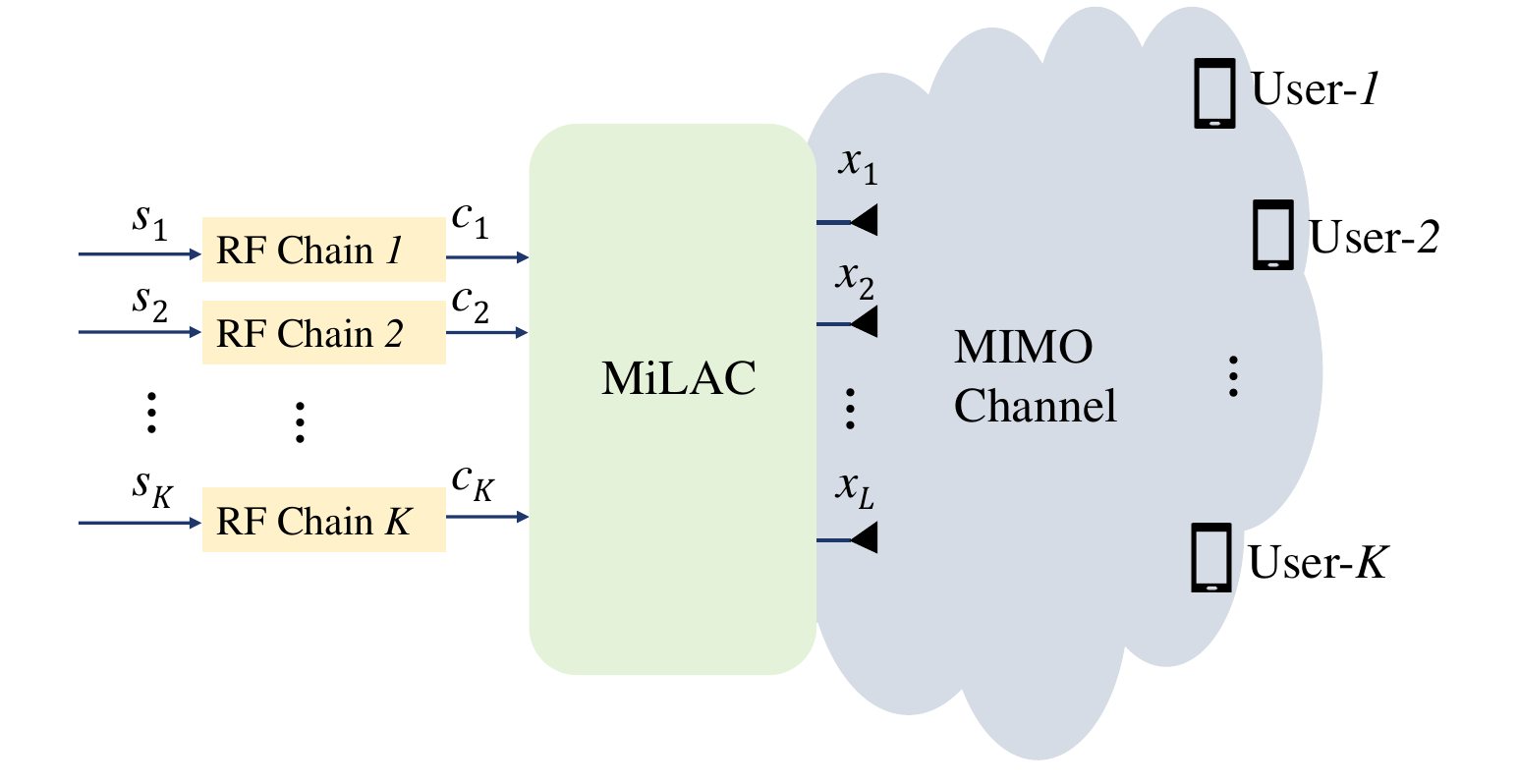}
		\caption{Multi-user MISO system with MiLAC at the BS. }
		\label{fig:sys-model}
        \vspace{-0.8cm}
	\end{figure}

	As shown in Fig.~\ref{fig:sys-model}, we consider a MU-MISO downlink communication system  where a base station (BS) equipped with a MiLAC-assisted antenna array of $ L $ transmit antennas and $K$ RF chains serves $ K $ single-antenna users.
	\subsubsection{Signal Model}
	Let $ \mathbf{s} = [s_1, \cdots, s_K]^T \in \mathbb{C}^{K \times 1}$ denote the information symbol vector, where $s_k \in \mathbb{C}$ is the symbol intended for user $k$. We assume that each symbol $s_k$ has zero mean and unit variance, i.e., $\mathbb E\{\mathbf s\mathbf s^H\}=\mathbf I$.	The source signal at the RF chains is denoted as $ \mathbf{c} = [c_1, \cdots, c_K]^T \in \mathbb{C}^{K \times 1} $, which incorporates the symbols and their allocated power, i.e.,
	\begin{equation}\label{eq:tx-c}
\mathbf{c}=\mathbf{P}^{1/2}\mathbf{s},
	\end{equation}
	where $ \mathbf{P}^{1/2} = \operatorname{diag} (\sqrt{p_1}, \cdots, \sqrt{p_K})$ is the square-root power allocation matrix. Here $ p_k $ denotes the transmit power assigned to symbol $s_k$ such that $ \sum_{k=1}^K p_k \leq P_\rmt $, with $P_\rmt$ being the total transmit power budget. 
    
    The MiLAC network further processes the RF-chain outputs through the beamforming matrix $\mathbf F\in\mathbb C^{L\times K}$, yielding the transmitted signal $ \mathbf x=\mathbf{F} \mathbf{c}$. Let $N\triangleq K+L$ denote the number of total ports at the MiLAC.  According to microwave network theory, the beamforming matrix $\mathbf F$ is determined by the admittance (Y-parameter) matrix $ \mathbf{Y} \in \mathbb{C}^{N \times N}$ of the MiLAC~\cite{nerini2025enabling}, which is given as
\begin{equation}\label{eq:beamforming}
		\mathbf{F} = \left[ \left({\mathbf{Y}}/{Y_0}+\mathbf{I}_{N}\right)^{-1} \right]_{K+1:K+L,1:K},
	\end{equation}
	where 
    $Y_0$ is the reference admittance.
    
    Let $\mathbf h_k\in\mathbb C^{L\times 1}$ denote the downlink channel vector from the BS to user $k$, assumed to be perfectly known at the BS. The received signal at user $ k $ is
	\begin{equation}\label{eq:user signal}
		l_k = \mathbf{h}_k^H \mathbf{x} + n_k = \mathbf{h}_k^H \mathbf{F} \mathbf{P}^{1/2}\mathbf s + n_k,
	\end{equation}
	where $n_k \sim \mathcal{CN}(0, \sigma_k^2)$ denotes the additive white Gaussian noise at user $k$. 
	
\subsubsection{MiLAC Model}
In this work, we consider a fully-connected MiLAC  architecture to enhance spectral efficiency. 
In the reconfigurable admittance network of a fully-connected MiLAC, each port $ v $ of the MiLAC is connected to ground through an admittance $ Y_{v,v} \in \mathbb{C}, \forall v \in \mathcal{V}=\{1,\cdots,N\}$, and it is interconnected to with every other port $ i $ through an admittance $ Y_{i,v} \in \mathbb{C}, \forall i \neq v$. Accordingly, the $(i,v)$-th entry of the MiLAC admittance matrix $\mathbf{Y}$ is given by
\begin{equation}
	[\mathbf{Y}]_{i,v} =  \left\{\begin{matrix}
	- Y_{i,v}	& i \neq v \\
	\sum_{j=1}^{N} Y_{j,v}	& i = v \\
	\end{matrix}\right..
\end{equation}
Assume the MiLAC is lossless and reciprocal, its  admittance components  are purely imaginary and can be expressed as $ Y_{i,v} = \rmj B_{i,v} $, where $\rmj$ denotes the imaginary unit and $ B_{i,v} \in \mathbb{R}$ denotes the susceptance associated with port pair $(i,v)$. Therefore, a lossless and reciprocal MiLAC should satisfy the following constraints
\begin{equation} \label{eq: Y parameter}
\mathbf{Y} = \rmj \mathbf{B}, \mathbf{B} = \mathbf{B}^T,
\end{equation}
where $ \mathbf{B} \in \mathbb{R}^{N \times N}$ is the susceptance matrix of the MiLAC. 

While the Y-parameters fully characterize the physical interconnections of the MiLAC network, the scattering (S-parameter) matrix is more commonly used to model the relationship between input and output signals. To be specific, the $ N $-port MiLAC network with admittance matrix $ \mathbf{Y}$ (here, equivalently, the susceptance matrix $\mathbf B$) can also be characterized via its scattering matrix $ \bm \Theta \in \mathbb{C}^{N\times N} $, which is related to $\mathbf B$ via the following relationship
\begin{equation}\label{eq:S-para}
	\bm \Theta = (\mathbf{I}_{N}+\rmj Z_0 \mathbf{B})^{-1}(\mathbf{I}_{N}-\rmj Z_0 \mathbf{B}),
\end{equation}
where $Z_0=Y_0^{-1}$ denotes the reference impedance. Given the MiLAC is lossless and reciprocal~\cite{pozar2021microwave}, constraint \eqref{eq:S-para} is mathematically equivalent to
\begin{equation}\label{eq:symuni}
	\bm \Theta^H\bm \Theta=\mathbf{I}_N, \bm \Theta=\bm \Theta^T.
\end{equation}

\subsubsection{Relationship between the Beamforming and Scattering Matrices} Next, we establish the relationship between the MiLAC beamforming matrix $\mathbf F$ and the scattering matrix $\bm\Theta$. Applying direct algebraic manipulation to  \eqref{eq:S-para} yields
\begin{align}\label{eq:ftheta}
\bm\Theta&=(\mathbf{I}_{N}+\rmj Z_0 \mathbf{B})^{-1}-(\mathbf{I}_{N}+\rmj Z_0 \mathbf{B})^{-1}\rmj Z_0 \mathbf B\nonumber\\
    &=(\mathbf{I}_{N}+\rmj Z_0 \mathbf{B})^{-1}+(\mathbf{I}_{N}+\rmj Z_0 \mathbf{B})^{-1}\nonumber\\&\quad\quad-(\mathbf{I}_{N}
    +\rmj Z_0 \mathbf{B})^{-1}-(\mathbf{I}_{N}+\rmj Z_0 \mathbf{B})^{-1}\rmj Z_0 \mathbf B\nonumber\\
    &=2(\mathbf{I}_{N}+\rmj Z_0 \mathbf{B})^{-1}-\mathbf I_N.
\end{align}
This symmetric unitary scattering matrix  $\bm\Theta $ can be further partitioned as
\begin{equation}\label{eq:partition Theta}
    \bm\Theta=\begin{bmatrix}
        \bm\Theta_{11} &\bm\Theta_{21}^T\\
        \bm\Theta_{21} &\bm\Theta_{22}
    \end{bmatrix}, 
\end{equation}
where $ \bm\Theta_{11}\in\mathbb C^{K\times K},\bm\Theta_{21}\in\mathbb C^{L\times K}$ and $\bm\Theta_{22}\in\mathbb C^{L\times L}.$ Combining \eqref{eq:beamforming}, \eqref{eq: Y parameter}, \eqref{eq:ftheta} and \eqref{eq:partition Theta}, we obtain
\begin{equation}\label{eq:F with Theta}
	\mathbf{F} = \frac{1}{2}\left[\bm \Theta +\mathbf I_N\right]_{K+1:K+L,1:K}=\frac{1}{2}\bm \Theta_{21}.
\end{equation}


The structural constraint of the beamforming matrix $\mathbf{F}$ in \eqref{eq:F with Theta} leads to the following fundamental limitation of MiLAC beamforming.

\begin{proposition}\label{pro}
    In a $K$-user multi-antenna system, lossless and reciprocal MiLAC-based transmit beamforming cannot achieve the same performance as fully digital beamforming except in two degenerate cases: (1) {a single-user scenario, i.e., $K=1$}; (2) {a multi-user scenario with mutually orthogonal channel directions}.
\end{proposition}
\textit{Proof:} In a $K$-user multi-antenna system, following the MiLAC-aided system model specified in \eqref{eq:tx-c}--\eqref{eq:F with Theta}, the transmitted signal at the BS is given by
\begin{equation}\label{eq:tx-x-milac}
    \mathbf x=\mathbf F\mathbf P^{1/2}\mathbf s=\frac{1}{2}\bm\Theta_{21}\mathbf P^{1/2}\mathbf s,
\end{equation}
with $\bm\Theta_{21}$ satisfying \eqref{eq:partition Theta} and \eqref{eq:symuni}. The transmit power constraint is $\frac{1}{4}\mathrm{Tr}(\bm\Theta_{21}^H\bm\Theta_{21}\mathbf P)\leq P_\rmt^{\text{MiLAC}}$.

In comparison, the transmitted signal for a classical fully digital beamformer is modeled as $ \mathbf x=\mathbf{W} \mathbf{s}$, where $\mathbf{W}$ is an arbitrary beamforming matrix satisfying the transmit power constraint, i.e., $\mathrm{Tr}(\mathbf W \mathbf W^H) \leq P_\rmt^{\text{digital}}$.  $\mathbf W$ can also be decomposed into 
\begin{equation}\label{eq:tx-W-digial}
    \mathbf W=\frac{1}{2}\mathbf D\mathbf P^{1/2},
\end{equation}
where $\mathbf D$ contains the normalized digital beamforming directions and $\frac{1}{2}\mathbf P^{1/2}$ is a square-root  diagonal power allocation matrix. 
The scaling factor $\frac{1}{2}$ is introduced  to enable a clear comparison with $\frac{1}{2}\bm\Theta_{21}\mathbf P^{1/2}$ for MiLAC as specified in \eqref{eq:tx-x-milac} \cite[Appendix]{nerini2025capacity}. The primary difference between MiLAC-aided and fully digital beamforming is thus reduced to a comparison between the matrix  $\bm\Theta_{21}$ for MiLAC and $\mathbf D$ for digital. Unlike the structural constrained $\bm\Theta_{21}$, $\mathbf D$ can be any arbitrary set of normalized beamforming vectors.

In the following, we compare the transmit signals of MiLAC-aided and fully digital beamforming under two conditions regarding the scattering submatrix  $\bm\Theta_{11}$:  $\bm\Theta_{11}\neq 0$ and  $\bm\Theta_{11}= 0$. 
\begin{itemize}
    \item Case 1: $\bm\Theta_{11}\neq \mathbf 0$.
     From the constraint $\bm \Theta^H\bm \Theta=\mathbf{I}_N$, the non-zero $\bm\Theta_{11}$ implies $\bm\Theta_{21}^H\bm\Theta_{21}\prec\mathbf I_K$. This, in turn, leads to a strict  transmit power reduction for the MiLAC architecture compared to the fully digital beamformer:
    \begin{equation}
        P_\rmt^{\text{MiLAC}}=\frac{1}{4}\mathrm{Tr}(\bm\Theta_{21}^H\bm\Theta_{21}\mathbf P)\overset{(a)}{<}\frac{1}{4}P_\rmt=P_\rmt^{\text{digital}},
    \end{equation}
    where step $(a)$ follows from $\bm\Theta_{21}^H\bm\Theta_{21}\prec\mathbf I_K$.
   Physically,  a non-zero $\bm\Theta_{11}$ means part of the input power is reflected (i.e., $\bm\Theta_{11}\neq \mathbf 0$), causing the MiLAC to radiate less power than the digital beamformer. This implies $ \bm\Theta_{21} \neq \mathbf D$ and MiLAC cannot achieve the digital design.
    \item Case 2: $\bm\Theta_{11}=\mathbf 0$. In this case, the unitary property $\bm \Theta^H\bm \Theta=\mathbf{I}_N$ leads to
    \begin{equation}
        \bm\Theta_{21}^H\bm\Theta_{21}=\mathbf I_K,
    \end{equation}
    indicating that the columns of the MiLAC-aided beamforming $\mathbf F=\frac{1}{2}\bm\Theta_{21}$ has orthogonal columns. However, a generic digital beamforming matrix $\mathbf D$ has \emph{non-orthogonal} columns. Therefore, unless $\mathbf D^H\mathbf D=\mathbf I_K$, MiLAC can not realize $\mathbf D$. This orthogonality condition aligns precisely with the two special cases stated in the proposition. (1) In single-user MIMO, the optimal digital beamformer is the singular vector (or matrix) of the channel, which is unitary by construction and can therefore be directly embedded into a unitary scattering matrix $\bm\Theta$. (2) in multi-user systems with mutually orthogonal channel directions, the optimal beamforming directions are therefore mutually orthogonal, and hence can be selected as the columns of a unitary scattering matrix, allowing MiLAC to reproduce the digital beamformer exactly.
\end{itemize}
In summary, the above two cases reveal fundamental limitations of the MiLAC: Case 1 incurs power loss at the transmit antennas, while Case 2 restricts the beamforming directions among users to be orthogonal. Consequently, the considered lossless and reciprocal MiLAC-aided beamforming  cannot, in general,  attain the same performance as fully digital beamforming.$\hfill \square$

\subsection{Problem Formulation}

To further assess the sum-rate performance of MiLAC-aided multi-user multi-antenna networks,  we now jointly optimize the power allocation matrix $\mathbf P$ and the scattering matrix $\bm\Theta$ to maximize the sum-rate.
Base on \eqref{eq:user signal}, the achievable SINR at user $ k $ to decode its intended symbol is given by
\begin{equation}\label{eq:rate}
	\gamma_k = \frac{p_k|\mathbf{h}_k^H\mathbf{f}_k|^2}{ \sum_{i=1,i\neq k}^Kp_i|\mathbf{h}_k^H\mathbf{f}_i|^2 + \sigma_k^2} ,
\end{equation}
where $ \mathbf{f}_k $ is the $ k $-th column of $ \mathbf{F}, \forall k \in \{1, \cdots, K\} $. Consequently, the sum-rate maximization problem is formulated as
\vspace{-0.5 cm}
\begin{subequations}\label{P1}
	\begin{align}
		\max_{\mathbf{P}, \bm \Theta} \quad &\sum_{k=1}^K \log(1+\gamma_k)\\
		\label{P1C1}\operatorname{s.t.}  \quad 	&  \mathbf{F} = \frac{1}{2}\left[\bm \Theta \right]_{K+1:K+L,1:K},\\
	\label{P1C2}	&\bm \Theta^H \bm \Theta = \mathbf{I}_N, \bm \Theta = \bm \Theta^T,\\
	\label{P1C3}	& \mathbf{P} =\operatorname{diag}(p_1,\cdots,p_K),\\
		\label{P1C4}& \mathrm{Tr}(\mathbf P) \leq P_\rmt.
	\end{align}
\end{subequations}
 Constraints \eqref{P1C1} and \eqref{P1C2} capture the physical limitations on the beamforming matrix $\mathbf F$ imposed by MiLAC, while constraints \eqref{P1C3} and \eqref{P1C4} are the source power budget. Problem \eqref{P1} is inherently non-convex and NP-hard due to the coupled non-convex fractional SINR expressions and the highly non-convex unitary constraint on the scattering matrix. In the next section, we develop an efficient optimization framework to address these challenges. 

\vspace{-0.2cm}
 
\section{Proposed Optimization Framework}
In this section, we first introduce a fractional programming (FP) method to transform problem \eqref{P1} into a more tractable form and then employ a block coordinate descent (BCD) framework to solve the transformed problem.  Specifically, we first apply the FP method to problem \eqref{P1}, leading to the following lemma.
\begin{lemma}
	By introducing auxiliary variables $ \bm \alpha=\{\alpha_1,\cdots,\alpha_K\} $, $ \bm \beta=\{\beta_1,\cdots,\beta_K\} $, problem \eqref{P1} is equivalently formulated as		\begin{subequations}\label{P2}
	\begin{align}		\max_{\bm\alpha,\bm\beta,\mathbf{P}, \bm \Theta} \quad &\sum_{k=1}^K f_k(\bm\alpha,\bm\beta,\mathbf{P}, \bm \Theta)\\
		\operatorname{s.t.}  \quad 	&  \eqref{P1C1},\eqref{P1C2},\eqref{P1C3},\eqref{P1C4},
	\end{align}
\end{subequations}
where
\begin{equation}
 \begin{aligned}
      f_k&\triangleq \log(1+\alpha_k)-\alpha_k+2\sqrt{1+\alpha_k}\sqrt{p_k}\Re\{\mathbf h_k^H\mathbf f_k \beta_k^*\}\\
     &-|\beta_k|^2\left(\sum_{i=1}^Kp_i|\mathbf h^H_k\mathbf f_i|^2+{\sigma_k^2}\right).\nonumber
 \end{aligned}
 \end{equation}
\end{lemma}
\textit{Proof:} The equivalence can be readily established by applying the FP framework proposed in~\cite{Shen2018fractional}.$\hfill \square$

Problem \eqref{P2} is a typical multi-block problem, which can be solved by the well-known BCD approach. To simply the design, we rewrite problem \eqref{P2} into a more explicit and compact form:
\begin{subequations}
	\begin{align}		\max_{\bm\alpha,\bm\beta,\mathbf{P}, \bm \Theta} \quad &2\Re\left\{\mathrm{Tr}\left(\mathbf F\mathbf{P}^{\frac{1}{2}}\bm\Sigma_1^H\mathbf H^H\right)\right\}-\mathrm{Tr}\left(\mathbf F\mathbf P\mathbf F^H\mathbf H\bm\Sigma_2\mathbf H^H \right)\\
		\operatorname{s.t.}  \quad 	&  \eqref{P1C1},\eqref{P1C2},\eqref{P1C3},\eqref{P1C4},
	\end{align}
\end{subequations}
where 
\begin{subequations}
	\begin{align*}
	\bm \Sigma_1 &\triangleq \operatorname{diag}(\sqrt{1+\alpha_1}\beta_1, \cdots, \sqrt{1+\alpha_K}\beta_K),\\
    \bm\Sigma_2&\triangleq\mathrm{diag}(|\beta_1|^2,\cdots,|\beta_K|^2), \quad \mathbf H\triangleq [\mathbf h_1,\cdots,\mathbf h_K].
	\end{align*}
\end{subequations}
In the following, we present the solution to each block of \eqref{P2} by fixing the other blocks.
\subsection{Auxiliary Variable Updates}
Given other variables fixed, the subproblems with respect to $\bm\alpha$ and $\bm\beta$ are unconstrained and convex. By checking their first-order optimality conditions, the solutions for $\bm\alpha, \bm\beta$ can be obtained as follows:
 \begin{align}\label{eq:auxilary update}
 \alpha^\star_k=\gamma_k,\,\,
\beta_k^\star=\frac{\sqrt{1+\alpha_k}\sqrt{p_k}
     \mathbf h_k^H\mathbf f_k}{\sum_{i=1}^Kp_i|\mathbf h_k^H\mathbf f_i|^2+{\sigma_k^2}}.
 \end{align}
\vspace{-0.8cm}
\subsection{Power Allocation Matrix Updates}

When $\bm\alpha,\bm\beta,\bm\Theta$ are fixed, the subproblem of \eqref{P2} with respect to $\mathbf P$ is given by
\begin{subequations}\label{P3:power-allo}
	\begin{align}
		\max_{\mathbf{P}} \quad &2\Re\left\{\mathrm{Tr}\left(\mathbf{P}^{1/2}\mathbf M\right)\right\}-\mathrm{Tr}\left(\mathbf P\mathbf N\right) \\
		\operatorname{s.t.}  \quad & \eqref{P1C3},\eqref{P1C4},
	\end{align}
\end{subequations}
where $\mathbf M\triangleq \bm\Sigma_1^H\mathbf H^H\mathbf F$ and $\mathbf N\triangleq\mathbf F^H\mathbf H\bm\Sigma_2\mathbf H^H\mathbf F $.
Define $\mathbf z\triangleq [z_1,\cdots,z_K]^T\in\mathbb R^{K\times 1}$ with $z_k=\sqrt{p_k}, \forall k\in\{1,\cdots,K\}$, problem \eqref{P3:power-allo} can be reformulated as
\begin{subequations}\label{P3:simpled}
	\begin{align}
		\max_{\mathbf z\succcurlyeq \mathbf 0} \quad &2\mathbf z^T\mathbf m-\mathbf z^T \mathrm{diag}(\mathbf n)\mathbf z \\
		\operatorname{s.t.}  \quad & \mathbf z^T\mathbf z \leq P_\rmt,
	\end{align}
\end{subequations}
where $\mathbf m\triangleq \mathrm{diag}(\Re \{\mathbf M\} ),\mathbf n\triangleq \mathrm{diag}(\mathbf N)$ and $\mathbf z\succcurlyeq\mathbf 0 $ denotes that the vector $\mathbf z$ is element-wise nonnegative. Problem \eqref{P3:simpled} is convex and can be solved via Lagrangian method. Its optimal solution is given by
\begin{equation}\label{update z}
    \mathbf z^\star=\left[\left(\mathrm{diag}(\mathbf n)+\mu\mathbf I_K\right)^{-1}\mathbf m\right]^{+},
\end{equation}
 where operator $[\mathbf z]^{+}\triangleq[[z_1]^{+},\cdots,[z_K]^{+}]^T$ with $[z_k]^{+}\triangleq \max\{z_k,0\}$. The scalar $\mu\geq 0$ is the optimal dual variable associated with the transmit power constraint and can be efficiently found via a  bisection search.

\subsection{Scattering Matrix Updates}
With $ \bm \alpha $, $ \bm \beta $, $ \mathbf{P} $ fixed, the subproblem to $ \bm \Theta $ is given by
\begin{subequations}\label{eq:MiLAC-opt-com}
	\begin{align}
		\max_{\bm \Theta} \quad & 2\Re\left\{\operatorname{tr}\left( \mathbf L_1^H\mathbf{F}\right)\right\} -\operatorname{tr}\left(\mathbf{F}\mathbf{P}\mathbf{F}^H\mathbf{H}\bm \Sigma_2 \mathbf{H}^H\right) \\
		\operatorname{s.t.}  \quad 	& \eqref{P1C1},\eqref{P1C2},
	\end{align}
\end{subequations}
where $\mathbf L_1\triangleq\mathbf{H}\bm \Sigma_1\mathbf{P}^{1/2}$. It can be observed that the matrix $ \mathbf{F} $ is defined as a scaled submatrix of $ \bm \Theta $ in the constraint \eqref{P1C1}, which introduces an additional coupling between the optimization variable $ \bm \Theta $ and the objective function. To address this issue, we introduce two selection matrices $ \mathbf{S}_1= [\mathbf{0}_{L\times K},\frac{1}{2}\mathbf{I}_L]\in \mathbb{R}^{L\times N} $ and $ \mathbf{S}_2 = [\mathbf{I}_K;\mathbf{0}_{L\times K}]\in \mathbb{R}^{N\times K} $, such that $ \mathbf{F} =  \mathbf{S}_1 \bm \Theta \mathbf{S}_2$. Hence, we can rewrite \eqref{eq:MiLAC-opt-com} as follows
\begin{equation}\label{eq:MiLAC-theta}
	\begin{aligned}
		\max_{\bm \Theta\in\mathcal M} \quad  2\Re\left\{\operatorname{tr}\left( \mathbf L_2^H \bm \Theta \right)\right\}-\operatorname{tr}\left(\bm \Theta\mathbf{X}_1\bm \Theta^H \mathbf{X}_2 \right) 
	\end{aligned}
\end{equation}
where $\mathbf L_2\triangleq \mathbf S_1^H\mathbf L_1\mathbf S_2^H,\mathbf{X}_1 = \mathbf{S}_2 \mathbf{P}\mathbf{S}_2^H,
		\mathbf{X}_2 =\mathbf{S}_1^H \mathbf{H}\bm \Sigma_2 \mathbf{H}^H\mathbf{S}_1$ and $\mathcal M\triangleq \{\bm\Theta\big| \bm \Theta^H \bm \Theta = \mathbf{I}_N, \bm \Theta = \bm \Theta^T\}$. Problem \eqref{eq:MiLAC-theta} can be solved by the successive linear approximation method proposed in~\cite{zhou2025joint}. Specially, considering that $\mathrm{Tr}(\bm\Theta\bm\Theta^H\mathbf X_2)=\mathrm{Tr}(\mathbf X_2)$ is satisfied at any feasible point, we can add $\lambda \mathrm{Tr}(\bm\Theta\bm\Theta^H\mathbf X_2)$ to the objective function in \eqref{eq:MiLAC-theta}, defined as
\begin{equation}
    h(\bm\Theta)=2\Re\left\{\operatorname{tr}\left( \mathbf L_2^H \bm \Theta \right)\right\}+\operatorname{tr}\left(\bm \Theta(\lambda\mathbf I_N-\mathbf{X}_1)\bm \Theta^H \mathbf{X}_2 \right),
\end{equation}
where $\lambda$ is a predefined constant. By setting $\lambda=\max_k\{p_k\}$, function $h(\bm\Theta)$ is guranteed to be convex with respect to $\bm\Theta$. Then, we are able to construct a first-order Taylor approximation of the original problem \eqref{eq:MiLAC-opt-com} at any feasible point $\bm\Theta^{[t]}$:
\begin{equation}\label{eq:MiLAC-theta-linear}
	\begin{aligned}
		\max_{\bm \Theta\in\mathcal M} ~~  \Re\left\{\operatorname{tr}\left( \bm \Theta^H\left(\mathbf{L}_2+\left(\lambda\mathbf{I}_N-\mathbf{X}_2\right)\bm \Theta^{[t]}\mathbf{X}_1 \right) \right)\right\}
	\end{aligned}
\end{equation}
Its optimal solution is
\begin{equation}\label{update Theta}
	\bm \Theta^{[t+1]} = \bm \Pi_{\mathcal M}\left(\mathbf{L}_2+\left(\lambda\mathbf{I}_N-\mathbf{X}_2\right)\bm \Theta^{[t]}\mathbf{X}_1 \right),
\end{equation}
where notation $\bm \Pi_{\mathcal M}(\mathbf X)$ denotes the symmetric unitary projection of square matrix $\mathbf X$ into the feasible set $\mathcal M$. The details of the projection can be found in \cite{fang2023low}. 

\setlength{\textfloatsep}{7pt}	
\begin{algorithm}[t!]
	\caption{Proposed algorithm to solve problem \eqref{P1}}
	\label{alg:proposed}
	\textbf{Initilization:} $ n\leftarrow 0 $, $\mathbf P^{[0]}, \bm\Theta^{[0]}$\;
	\Repeat{The objective value in \eqref{P1} converges}{ 
		$ n\leftarrow n+1 $\;
		Update $ \alpha_{k}^{[n]} $ and $ \beta_{k}^{[n]} $ by \eqref{eq:auxilary update}, $ \forall k\in\mathcal{K} $\;
		 Find optimal $\mu^{[n]}$ via a bisection search\;
         Update $\mathbf z^{[n]}$ using \eqref{update z} and $\mathbf P^{[n]}=[\mathrm{diag}(\mathbf z^{[n]})]^2$\;
		Let $\bm\Phi^{[0]}=\bm\Theta^{[n-1]}$ and $\lambda=\max_k\{p_k^{[n]}\}$\;
		\For{$t=1:I_1$}{ 
        
			$ \bm\Phi^{[t]}= \bm \Pi_{\mathcal M}\left(\mathbf{L}_2+\left(\lambda\mathbf{I}_N-\mathbf{X}_2\right)\bm \Phi^{[t-1]}\mathbf{X}_1 \right)$\;
		}
        Update $\bm\Theta^{[n]}=\bm\Phi^{[I_1]}$\;
        Update $\mathbf{F}^{[n]} = \frac{1}{2}\left[\bm \Theta^{[n]} \right]_{K+1:K+L,1:K}$;
	}
\end{algorithm}
\vspace{-0.4cm}
\subsection{Overall Algorithm Development}
For clarity, the complete procedure for solving problem \eqref{P1} is summarized in Algorithm \ref{alg:proposed}. The algorithm starts from any feasible non-zero initialization $\bm\Theta^{[0]}$ and $\mathbf P^{[0]}$. At each outer iteration, the auxiliary variables $\{\alpha_k^{[n]},\beta_k^{[n]}\}$ are first updated in closed form according to \eqref{eq:auxilary update}. Then, the power-allocation subproblem is solved by computing the optimal dual variable $\mu^{[n]}$ via a bisection search, followed by updating the optimal auxiliary variable $\mathbf z^{[n]}$ and the power matrix $\mathbf P^{[n]}$ using \eqref{update z}. Next, the scattering matrix is refined through $I_1$ inner iterations, where each update applies \eqref{update Theta}. These steps are repeated until the objective value of problem \eqref{P1} converges. This iterative procedure is guaranteed to produce a feasible sequence and yields a monotonically improving objective value. 
\vspace{-0.4cm}
\subsection{Convergence and Complexity Analysis}

The convergence analysis of Algorithm~\ref{alg:proposed} follows the procedure in \cite{zhou2025joint} and is omitted here for brevity. Its convergence behavior will be verified in the following Section \ref{Sec:numerical}.
For Algorithm~\ref{alg:proposed}, the computational complexity is dominated by the update of the scattering matrix (line 9), which has an order of $\mathcal{O}\big((K+L)^3\big)$. Consequently, the overall complexity of Algorithm~\ref{alg:proposed} is $\mathcal{O}\left(I_2 I_1 (K+L)^3\right)$, where $I_2$ and $I_1$ denote the number of iterations required for the outer loop and the inner loop, respectively.

\vspace{-0.4cm}
\section{Simulation Results}\label{Sec:numerical}
In this section, we present numerical results to validate the theoretical analysis and demonstrate the effectiveness of the proposed algorithm. We consider a MISO broadcast channel (BC) under two different channel models. We first examine the classical Rayleigh fading scenario, where the channel of user $k$ is generated i.i.d as $\mathbf h_k\sim \mathcal {CN}(\mathbf 0, \mathbf I_L)$, and the noise variance is fixed at $\sigma_k^2=1$. Under this setting, the transmit SNR, defined as $\mathrm{SNR}=P_\rmt/\sigma_k^2$, is numerically equal to the transmit power. The number of inner iterations for updating $\bm\Theta$ is set to $I_1=50$, and the convergence tolerance for Algorithm \ref{alg:proposed} is chosen as $10^{-4}$. In addition, to further evaluate Proposition \ref{pro}, we also consider synthetically generated orthogonal channel realizations, constructed by taking the product of the left singular matrix and the diagonal singular-value matrix of the Rayleigh channel. All simulation results are averaged over 100 independent channel realizations.

\begin{figure}[t]
	\centering
\includegraphics[width=0.5\linewidth]{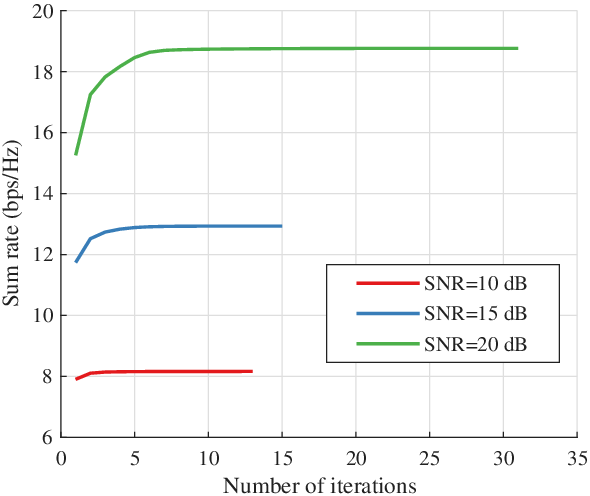}
	\caption{Convergence behavior of the proposed algorithm. }
	\label{fig:convergence}
\end{figure} 

Fig.\ \ref{fig:convergence} illustrates the convergence behavior of the proposed Algorithm \ref{alg:proposed} under different transmit SNRs. It is observed that, in all cases, the algorithm exhibits a monotonic increase in the objective value and converges rapidly within a limited number of iterations, thereby validating the effectiveness and stability of the proposed optimization framework.

\begin{figure}[t]
    \centering
    \begin{subfigure}[t]{0.48\linewidth}
        \centering
        \includegraphics[width=\linewidth]{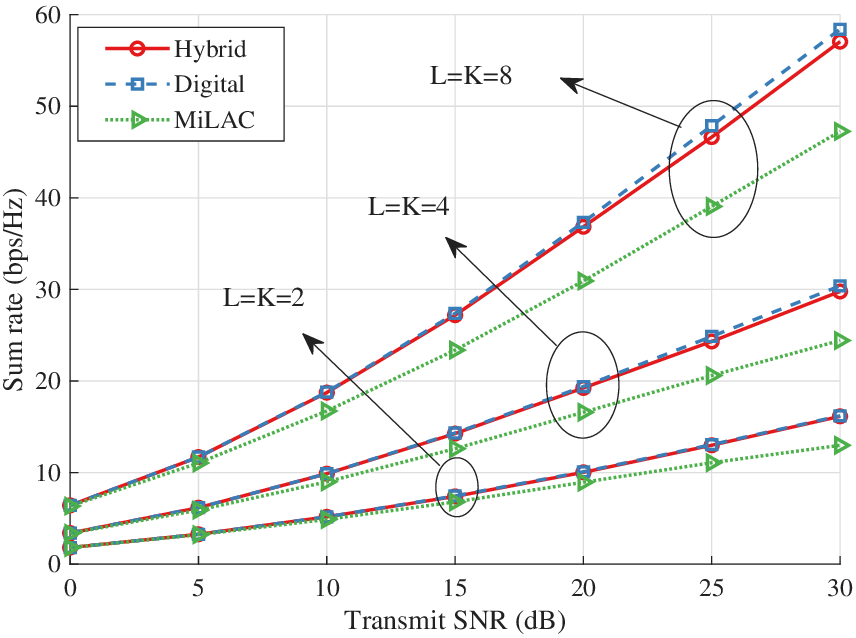}
        \caption{Rayleigh channel.}
        \label{fig:SR_vs_Pt_ray}
    \end{subfigure}
    \hfill
    \begin{subfigure}[t]{0.48\linewidth}
        \centering
        \includegraphics[width=\linewidth]{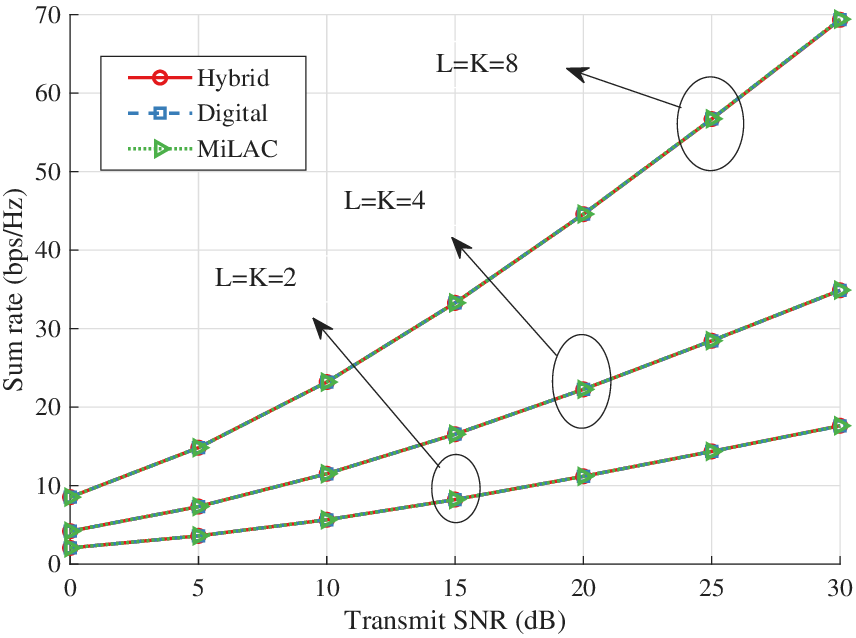}
        \caption{Orthogonal channel.}
        \label{fig:SR_vs_Pt_ort}
    \end{subfigure}
    \caption{Sum rate versus transmit SNR under two channel models.}
    \label{fig:Sumrate_vs_Pt}
\end{figure}

Fig.\ \ref{fig:Sumrate_vs_Pt} illustrates the sum-rate performance versus the transmit SNR under two channel models, compared against two conventional architectures: fully digital beamforming and hybrid digital–analog beamforming with a fully-connected structure \cite{yu2016alternating}. In Fig.\ \ref{fig:SR_vs_Pt_ray}, digital beamforming achieves slightly higher performance than hybrid beamforming, while both provide a substantial gain over MiLAC beamforming. The performance loss of MiLAC primarily stems from the physical constraints imposed by the lossless and reciprocal MiLAC network, together with the absence of digital beamforming capabilities. Consequently, multi-user interference cannot be effectively managed using only the MiLAC network, and the performance gap widens as the network load or the transmit SNR increases. Fig.\ \ref{fig:SR_vs_Pt_ort} further validates Proposition~\ref{pro}, showing that MiLAC matches the performance of digital beamforming when the desired beamforming directions are orthogonal.

\begin{figure}[t]
    \centering
    \begin{subfigure}[t]{0.48\linewidth}
        \centering
        \includegraphics[width=\linewidth]{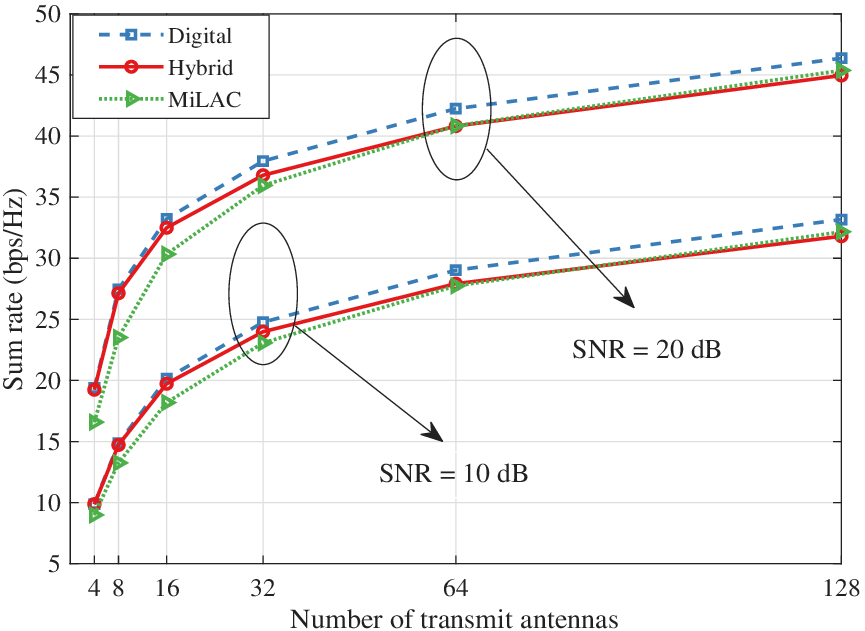}
        \caption{Rayleigh channel.}
        \label{fig:SR_vs_L_ray}
    \end{subfigure}
    \hfill
    \begin{subfigure}[t]{0.48\linewidth}
        \centering
        \includegraphics[width=\linewidth]{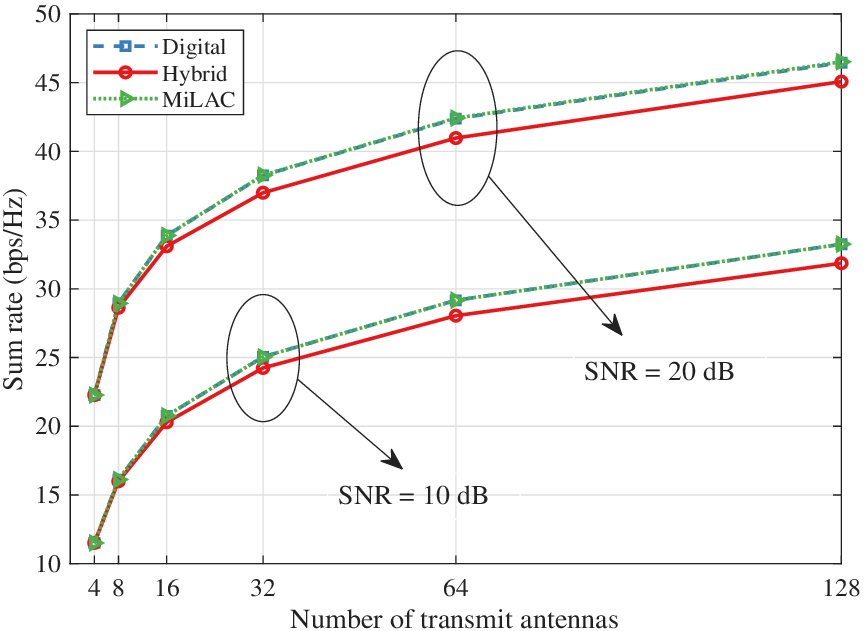}
        \caption{Orthogonal channel.}
        \label{fig:SR_vs_L_ort}
    \end{subfigure}
    \caption{Sum rate versus transmit SNR under two channel models.}
    \label{fig:Sumrate_vs_L}
\end{figure}

Fig.\ \ref{fig:Sumrate_vs_L} illustrates the sum-rate performance as the number of transmit antennas increases. In Fig.\ \ref{fig:SR_vs_L_ray}, the performance loss of MiLAC over fully digital beamforming initially grows and then decreases, whereas the performance loss of hybrid beamforming increases monotonically. This behavior occurs because the channels become asymptotically orthogonal as the transmit array size grows, which substantially enhances the performance of MiLAC. Notably, the MiLAC architecture outperforms hybrid beamforming when the number of transmit antennas reaches $128$, highlighting its strong potential for extremely large-scale MIMO systems. Fig.\ \ref{fig:SR_vs_L_ort} further validates Proposition~\ref{pro}, showing an almost perfect match with digital beamforming.

\vspace{-0.4cm}
\section{Conclusion}

This work analyzed the performance of a fully-connected, lossless, and reciprocal MiLAC architecture for downlink MU-MISO systems. We proved analytically that a lossless and reciprocal MiLAC cannot attain the same performance as digital beamforming in a general MU-MISO network, and proposed an efficient algorithm for joint RF-chain power allocation and MiLAC configuration. Simulation results showed that while MiLAC suffers performance loss at moderate transmit array sizes, the gap narrows as the array size grows, and MiLAC can even outperform hybrid beamforming at large array scales. Moreover, MiLAC matches digital beamforming under asymptotically orthogonal channels, validating our theoretical analysis. These findings highlight MiLAC as a promising architecture for future extremely large-scale MIMO systems.

\bibliographystyle{IEEEtran}  
\bibliography{reference}

\end{document}